\def\figcond{1}     
\def\figs#1#2#3#4#5#6#7{%
\ifnum\figcond>0
  \centerline{\hbox{\hsize=6in\hss\vbox to #5in{\vss%
  \vskip #6in\centerline{\hskip #4in\epsfig{file=#1,width=#3in}}
  \vss}\hss}}
\else
  \centerline{\hbox{\hsize=6in\vrule\hss\vbox to #5in{\hrule\vss%
  \centerline{Figures or Hardcopys available from %
              koepf@josiah.tau.ac.il}%
  \vss\hrule}\hss\vrule}}
\fi
\vskip 0.15in
\begin{quote}
\noindent Fig.~#2: #7
\end{quote}
\vskip -0.15in
}
\ifnum\figcond>0
\documentstyle[12pt,epsfig]{article}
\else
\documentstyle[12pt]{article}
\fi
\newcommand{\beq}{\begin{equation}}
\newcommand{\eeq}{\end{equation}}
\newcommand{\beqa}{\begin{eqnarray}}
\newcommand{\eeqa}{\end{eqnarray}}

\begin{document}

\hfill{TAUP-2298-95, hep-ph/9510452}

\smallskip

\begin{center}

{\large {\bf THE NUCLEON'S ANTIQUARK SEA:}}\\
{\large {\bf A VIRTUAL MESON CLOUD OR GLUONS}{\LARGE ?}}

\vspace{0.4cm}

W. Koepf and L.L. Frankfurt\\
{\it School of Physics and Astronomy, Tel Aviv University,
69978 Ramat Aviv, Israel}
\vspace{0.4cm}

M. Strikman\\
{\it Pennsylvania State University, University Park, PA 16802, USA}

\vspace{0.3cm}

\end{center}

\baselineskip 10pt

\begin{quote}
We study the possible contribution of the nucleon's virtual meson cloud to the
sea quark distribution as observed in deep inelastic lepton scattering.  We
adjust the meson-nucleon cut-offs to the large-$x$ tails of the antiquark
distributions, find qualitatively different behavior in the flavor singlet and
non-singlet channels and study the scale dependence of our results.  We
demonstrate that, within convolution models, to reproduce the sea quark
distribution the relevant pion momenta should be around 0.8 GeV.
\end{quote}

\baselineskip 14pt

\vspace{0.3cm}
\noindent
{\bf 1~~~The Nucleon's Meson Cloud and Deep Inelastic Lepton Scattering}
\vspace{0.3cm}

\noindent
In this realm, the nucleon is pictured as being part of the time a
bare core and part of the time a baryon with one meson "in the air",
\begin{equation}
|N\rangle = \sqrt{Z} \, |N_0\rangle + \sum_M g_{MNB} |BM\rangle \ .
\end{equation}
The wave function renormalization factor, $\sqrt{Z}$, only affects the core
because physical (renormalized) and not bare couplings were used [1].
Thus, we keep the conventional value for the $\pi NN$ coupling constant, which
follows unambigously from dispersion relations for the $\pi N$ amplitude. In
this point our approach differs from other prescriptions [2] where a
renormalization of the $\pi NN$ coupling constant has been introduced.
The meson cloud contribution to the nucleon's antiquark sea can be written as,
\begin{equation}
x \, \bar q_N(x,Q^2) = \sum_{M,B} \alpha_{MB}^q
\int^1_x dy \, f_{MB} (y) \,
{x \over y} \, \bar q_M\!\left({x \over y},Q^2\right) \ ,
\end{equation}
where $\alpha_{MB}^q$ are spin-flavor $SU(6)$ Clebsch-Gordan factors,
$x \bar q_M (x,Q^2)$ is the meson's valence antiquark distribution [3]
fit to Drell-Yan pair production experiments and
\begin{equation}
f_{MB}(y) = {g_{MNB}^2 \over 16\pi^2} \, y \, \int_{-\infty}^{t_{min}}
             dt \, {{\cal I}(t,m_N,m_B) \over (t-m_M^2)^2} \,
             F^2_{MNB}(t)
\end{equation}
is the meson's light-cone distribution in the nucleon's cloud.
Here, $t_{min} = m_N^2 y - m_B^2 y / (1-y)$ and
\begin{equation}
{\cal I}(t,m_N,m_B) = \left\{
\begin{array}{ll}
    -t+(m_B-m_N)^2
    & \qquad \mbox{for $B \in$ {\bf 8}} \\
    \noalign{\medskip}
    {\big((m_B+m_N)^2-t\big)^2 \big((m_B-m_N)^2-t\big)
    \over 12 m_N^2 m_B^2 }
    & \qquad \mbox{for $B \in$ {\bf 10}} \ .
\end{array} \right.
\end{equation}
The only unknown quantity in the above equations is
the form factor, which governs the emission of an
off-mass-shell meson, and we parametrize it in exponential form:
\begin{equation}
F_{MNB}(t) = e^{(t-m_M^2)/\Lambda^2_{MNB}} \ .
\end{equation}

\vspace{0.6cm}
\noindent {\bf 2~~~The Nucleon's Antiquark Sea}
\vspace{0.3cm}

\noindent
We adjust the various cut-offs
to the nucleon's strange quark content,
$x \bar s$, and
to the $SU(3)$ and $SU(2)$ flavor breaking components of the nucleon's
antiquark sea, $x \bar q_8 =  x(\bar u + \bar d - 2 \bar s)$ and
$x \bar q_3 = x(\bar d - \bar u)$. This yields
\begin{eqnarray}
\Lambda_{\pi NN}      & \approx & ~1000~\mbox{MeV}
\nonumber\\
\Lambda_{\pi N\Delta} & \approx & ~~800~\mbox{MeV}
\\
\Lambda_{KNY}         & \approx & ~1200~\mbox{MeV}
\nonumber
\end{eqnarray}
at a scale of $Q^2=1$ GeV$^2$. Corresponding results are depicted in Fig.~1.

\figs{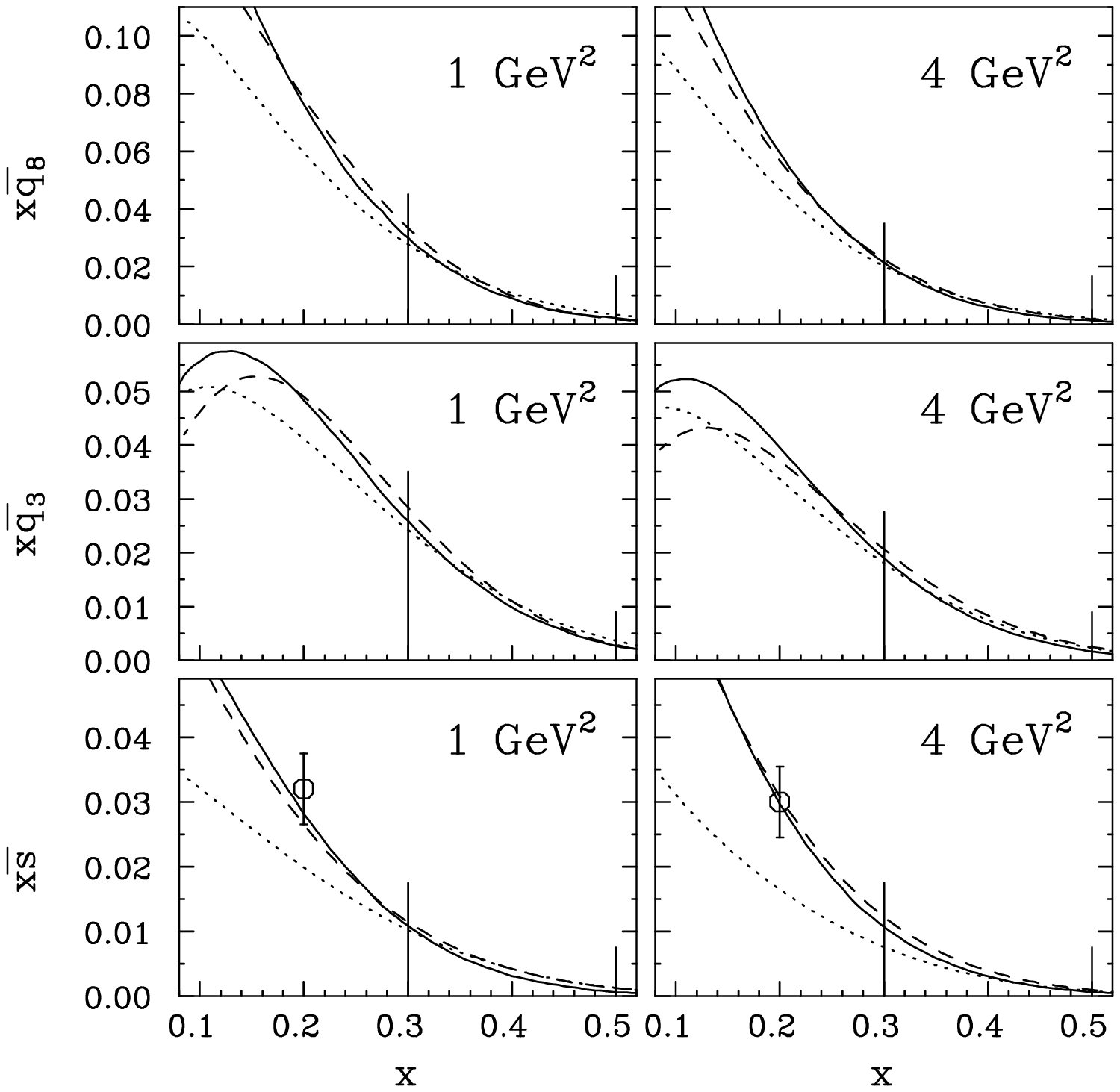}{1}{5.25}{-0.55}{4.0}{0.9}{The various components of
the nucleon's antiquark sea. The solid and dashed curves show the
MRS(A) [4] and CTEQ3M [5] parametrizations, the dotted
line refers to our meson cloud calculation and the data points are from
the CCFR collaboration [6].}

The agreement in the flavor breaking channels
extends to smaller $x$-values than those actually
considered in our fits ($0.3 \le x \le 0.5$) and it remains
satisfactory when the four-momentum transfer is increased.
The situation is very different for the $x \bar s$ component, which is
predominantly flavor singlet. Here, we only find reasonable agreement with the
large-$x$ tail and the deviations between our fit and the data grow rapidly
with increasing $Q^2$.

\vspace{0.6cm}
\noindent {\bf 3~~~Discussion}
\vspace{0.3cm}

\noindent
The vertices in Eq.~(6) are
softer than what is used in most meson-exchange models of the $NN$
interaction [7] and they are harder than those inferred from considerations
of integrated quantities (sum rules) in DIS [8]. The fact that the
$\pi N\Delta$ vertex is considerably softer than the $\pi NN$ vertex
indicates significant $SU(6)$ breaking in the $N$/$\Delta$ system.
A quantitative measure of the importance of the mesonic component
is the "number of mesons in the air",
which is $n_\pi = 0.66$ and $n_K = 0.10$.

In Fig.~2, we analyze which meson
virtualities and loop momenta yield the dominant share to the convolution
integral of Eq.~(2). We find that, typically, the mesons are highly
virtual, $-t \approx 0.4$ GeV$^2$, and the integrals are dominated by large
pion momenta of the order of $|{\bf k}| \approx 0.8$ GeV. Thus, the periphery
of the nucleon, where the notion of a pion cloud is well defined theoretically,
gives an insignificant contribution to the quark sea within a nucleon.

\figs{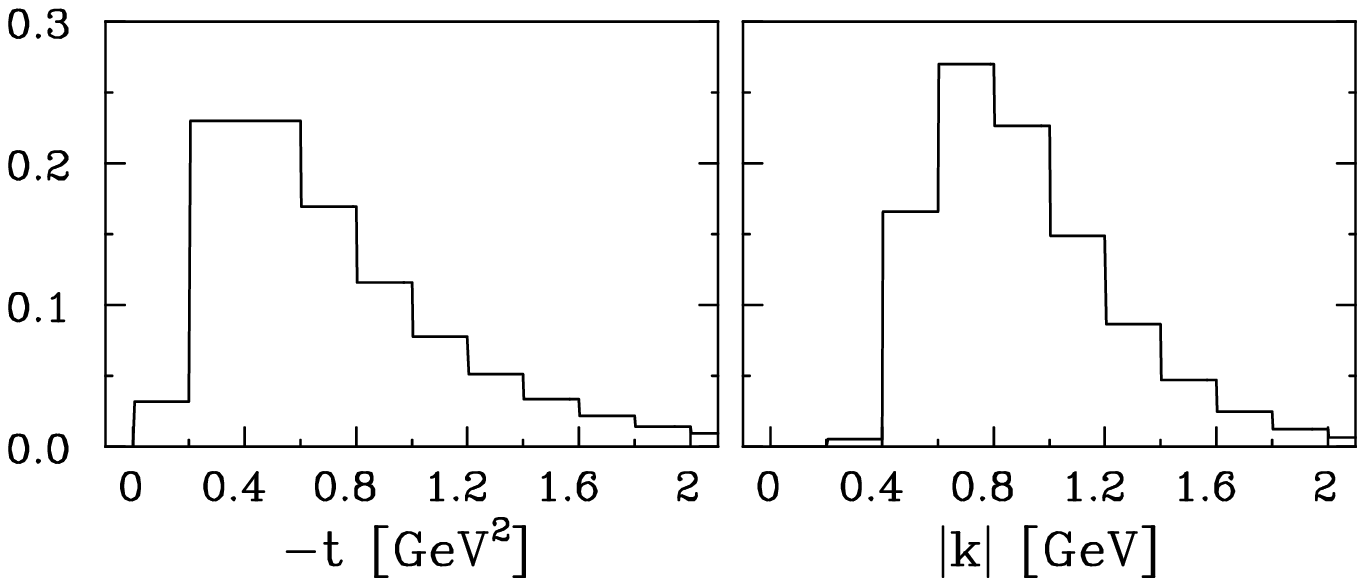}{2}{5.25}{0.0}{1.5}{0.5}{The different relative
contributions to the convolution integral of Eq.~(2) for the $N \to
N\pi$ sub-process and for a typical $x$-value of $x=0.3$.}

If we evaluate the meson cloud contribution to the violation of the Gottfried
sum rule,
\begin{equation}
\int^1_0 dx \, {F_2^{\mu p}(x,Q^2) - F_2^{\mu n}(x,Q^2) \over x}
= {1 \over 3} - {2 \over 3} \Delta_{G} =
  {1 \over 3} - {2 \over 3} \int^1_0 dx \,
                     [\bar d(x,Q^2) - \bar u(x,Q^2)] \ ,
\end{equation}
we find
a value of $\Delta_G = 0.17$ at $Q^2=4$ GeV$^2$. This agrees
with the quantity given by the NMC collaboration of
$\Delta_G^{exp} = 0.148 \pm 0.039$ [9]. However, the dominant
contribution to $\Delta_G$ comes from the small-$x$ region
where the meson cloud convolution picture is questionable due to shadowing
effects. It was argued in Ref.~[10] that $\Delta_G^{exp}$ could already
be saturated by the region $0 < x < 0.02$ through $A_2$ Reggeon exchange.
This indicates that the violation of the Gottfried
sum rule is not necessarily a mesonic effect.

The description of the {\em
flavor non-singlet} share of the nucleon's antiquark distributions is
quite satisfactory for $x$-values larger than about 0.2, and the
quality of the agreement is independent on the scale $Q^2$. This indicates
that the evolution of the partonic distributions in the framework of
pQCD and the convolution picture are compatible with
each other in the non-singlet channels.
The {\em flavor singlet} component, on the other hand, is quite
poorly described through the mesonic cloud, even at a small
scale of $Q^2=1$ GeV$^2$.  We are, in fact, only able to attribute
the large-$x$ tail of the $x \bar s$ distribution to the nucleon's
virtual meson cloud, and the deviations from the data-based
parametrizations grow rapidly with increasing $Q^2$. This indicates
that, in the singlet channel, other degrees of freedom, most notably
gluon splitting into $q\bar q$ pairs, are relevant, even at moderate
$x$ and small values of the four-momentum transfer. This hints that it is not
possible to attribute the entire sea quark distribution in the nucleon to
its mesonic cloud.

In DIS at high energies, the virtual photon converts
into a hadronic $q \bar q$ state at a distance $l \approx 1 / 2 m_N x$
before the target.
If this coherence length is larger than the dimension of the target,
$l \geq 2 \, \langle r_T \rangle$, not the virtual photon is probing the
target but this $q \bar q$ state and the na\"\i ve impulse approximation
picture is no longer applicable.
As the relevant meson loop momenta are $|{\bf k}|
\approx 0.8$ GeV, significant distances
are of the order of $\langle r_{MN}
\rangle \approx 0.25$ fm and the shadowing condition
is already satisfied at $x \leq 0.2$. This underlines our conclusion
that at small $x$ the meson cloud picture is not adequate and different
degrees of freedom are dominant.

\vspace{0.6cm}
\noindent {\bf 4~~~Summary}
\vspace{0.3cm}

\noindent
We critically discussed the analysis of the
contribution of the nucleon's virtual meson cloud to DIS.
The meson-nucleon cut-offs that we
determine are softer than those employed in most effective $NN$ potentials.
However, the meson loop momenta that are probed in DIS, $|{\bf k}| \approx 0.8$
GeV, are very different from those relevant for low-energy nuclear
physics phenomena.
While the agreement of our calculations
with the data-based parametrizations is satisfactory and scale
independent in the flavor breaking channels,
the flavor singlet component
is quite poorly described in the convolution picture.
This stresses the importance of gluonic degrees of freedom, even
at such a low scale as $Q^2=1$ GeV$^2$. For further details see Ref.~[11].

\vspace{0.6cm}
\noindent {\bf Acknowledgements}
\vspace{0.3cm}

\noindent
This work was supported in part by the
Israel-USA Binational Science Foundation Grant No.~9200126, by the
MINERVA Foundation of the Federal Republic of Germany, and by the
U.S.~Department of Energy under Contract No.~DE-FG02-93ER40771.

\vspace{0.6cm}
\noindent{\bf References}

\begin{itemize}
\parskip=0pt

\item[1.] W. Melnitchouk and A.W. Thomas, {\it Phys. Rev.\/} {\bf D47}, 3794
(1993).

\item[2.] V.R. Zoller, {\it Z. Phys.\/} {\bf C53}, 443 (1992);
A. Szczurek, and J. Speth, {\it Nucl. Phys.\/} {\bf A555}, 249 (1993).

\item[3.] M. Gl\"uck E. Reya and A. Vogt, {\it Z. Phys.\/} {\bf C53},
651 (1992); P.J. Sutton, A.D. Martin, R.G. Roberts and W.J. Stirling,
{\it Phys. Rev.\/} {\bf D45}, 2349 (1992).

\item[4.] A.D. Martin, W.J. Stirling and R.G. Roberts, {\it Phys. Rev.\/}
{\bf D50}, 6734 (1994).

\item[5.] H.L. Lai {\it et al.\/} (CTEQ Collaboration), {\it Phys. Rev.\/} {\bf
D51},
4763 (1995).

\item[6.] A.O. Bazarko {\it et al.\/} (CCFR Collaboration), {\it Z. Phys.\/}
{\bf C65},
189 (1995).

\item[7.] R. Machleidt, K. Holinde and Ch. Elster, {\it Phys. Rep.\/} {\bf
149},
1 (1987).

\item[8.] A. W. Thomas, {\it Phys. Lett.\/} {\bf 126B}, 97 (1983).

\item[9.] P. Amaudruz {\it et al.\/}, {\it Phys. Rev. Lett.\/} {\bf 66}, 2712
(1991)
and {\it Phys. Rev.\/} {\bf D50}, R1 (1994).

\item[10.] M.~Strikman, Proceedings of the XXVI International Conference on
High Energy Physics, Vol.~I, Dallas, USA, 1992,
{\it AIP Conference Proceedings\/} {\bf 272}, 806 (1992).

\item[11.] W. Koepf {\it et al.\/}, Preprint TAUP-2273-95 and
hep-ph/9507218 (1995).

\end{itemize}

\end{document}